\documentclass[10pt]{article}

\usepackage{geometry}
\usepackage{amsfonts}
\usepackage{amsmath}
\usepackage{amssymb}
\usepackage{amsthm}
\usepackage{algorithm}
\usepackage[noend]{algpseudocode}
\usepackage{booktabs}
\usepackage{subfig}
\usepackage{url}
\usepackage{empheq}
\usepackage{color}
\usepackage{graphicx}
\usepackage[colorlinks=true]{hyperref}
\usepackage[font=small]{caption}

\newcommand{\ip}[2]{\langle{#1},{#2}\rangle}

\newcommand{\cost}{\operatorname{cost}}
\newcommand{\avg}{\operatorname{average}}

\renewcommand{\vec}[1]{{\bf #1}}

\begin{document}
\title{Mapping constrained optimization problems to
  quantum annealing with application to fault diagnosis}
  
\author{Z. Bian\,$^{1}$, F. Chudak\,$^{1}$, R. Israel\,$^{1}$, B. Lackey\,$^{2}$, W. G. Macready\,$^{1}$, and A. Roy\,$^1$ \\
$^{1}$D-Wave Systems, Burnaby, BC, Canada \\
$^{2}$Joint Institute for Quantum Information and Computer Science, \\
University of Maryland, College Park, MD, USA} \date{\today}

\maketitle

\begin{abstract}
Current quantum annealing (QA) hardware suffers from practical limitations such as finite temperature, sparse connectivity, small qubit numbers, and control error. We propose new algorithms for mapping boolean constraint satisfaction problems (CSPs) onto QA hardware mitigating these limitations. In particular we develop a new embedding algorithm for mapping a CSP onto a hardware Ising model with a fixed sparse set of interactions, and propose two new decomposition algorithms for solving problems too large to map directly into hardware.

The mapping technique is locally-structured, as hardware compatible Ising models are generated for each problem constraint, and variables appearing in different constraints are chained together using ferromagnetic couplings. In contrast, global embedding techniques generate a hardware independent Ising model for all the constraints, and then use a minor-embedding algorithm to generate a hardware compatible Ising model.  We give an example of a class of CSPs for which the scaling performance of D-Wave's QA hardware using the local mapping technique is significantly better than global embedding. 

We validate the approach by applying D-Wave's hardware to circuit-based fault-diagnosis. For circuits that embed directly, we find that the hardware is typically able to find \emph{all} solutions from a min-fault diagnosis set of size $N$ using $1000N$ samples, using an annealing rate that is $25$ times faster than a leading SAT-based sampling method. Further, we apply decomposition algorithms to find min-cardinality faults for circuits that are up to 5 times larger than can be solved directly on current hardware. 
\end{abstract}

\section{Introduction}
\label{sec:background}

In the search for ever faster computational substrates recent
 attention has turned to devices manifesting quantum effects. Since it has long been realized that computational speedups may be obtained through exploitation of quantum resources, the construction of devices realizing these speedups is an active research area. Currently, the largest scale computing devices using quantum resources are based on physical realizations of quantum annealing. Quantum annealing (QA) is an optimization heuristic sharing much in common with simulated annealing, but which utilizes quantum, rather than thermal, fluctuations to foster exploration through a search space \cite{Finilla94,Kadowaki98,Farhi2000}.

QA hardware relies on an equivalence between a physical quantum model
and a useful computational problem. The low energy physics of the
D-Wave QA device \cite{Berkley2010,Dickson2013,Harris2010,Johnson2011} is well captured by a time-dependent Hamiltonian of
the form $\vec{H}(t) = A(t)\vec{H}_0 + B(t) \vec{H}_P$ where $\vec{H}_0 =
\sum_{i\in V(G)}\vec{\sigma}^x_i$ includes off-diagonal quantum effects, and where
$\vec{H}_P = \sum_{i\in V(G)}h_i \vec{\sigma}^z_i + \sum_{(i,j)\in
E(G)}J_{i,j} \vec{\sigma}^z_i\vec{\sigma}_j^z$ is used to
encode a classical Ising optimization problem of the form
\begin{equation} 
\min_{\vec{s}} E(\vec{s}) \equiv \min_{\vec{s}} \Bigl\{ \sum_{i\in V(G)} h_i s_i + \sum_{(i,j)\in
E(G)}J_{i,j}s_i s_j  \Bigr\}. \label{eq:Ising}
\end{equation} 
On the D-Wave device the connectivity between the binary variables
$s_i\in\{-1,+1\}$ is described by a fixed sparse graph $G=(V,E)$. The weights
$\vec{J} \equiv \{J_{i,j}\}_{(i,j)\in E(G)}$, and the linear biases
$\vec{h} \equiv \{h_i\}_{i\in V(G)}$ are programmable by the user. The
$A(t)$ and $B(t)$ functions have units of energy, and satisfy
$B(t=0)=0$ and $A(t=\tau)=0$ so that as time advances from $t=0$ to
$t=\tau$ the Hamiltonian $\vec{H}(t)$ is annealed to a purely classical
form. Thus, the easily prepared ground state of $\vec{H}(0)=\vec{H}_0$
evolves to a low energy state of $\vec{H}(\tau)=\vec{H}_P$, and
measurements at time $\tau$ yield low energy states of the classical
Ising objective Eq.~\eqref{eq:Ising}. Theory has shown that if the
time evolution is sufficiently slow, \textit{i.e.} $\tau$ is
sufficiently large, then with high probability the global minimizer of
$E(\vec{s})$ can be obtained.

Physical constraints on current hardware platforms \cite{Bunyk2014} impact this theoretical efficacy of QA.  \cite{Bian14} has noted the following
issues that are detrimental to performance:
\begin{enumerate}
\item \textit{Limited precision/control error on parameters $\vec{h}$/$\vec{J}$}:
  problems are not represented exactly in hardware, but are subject to
  small, but noticeable, time-dependent and time-independent additive
  noise.
\item \textit{Limited range on $\vec{h}$/$\vec{J}$ bounds the range of all
  parameters relative to thermal scales $k_bT$}: thus very low
  effective temperatures which are needed for optimization when there
  are many first excited states are unavailable.
\item \textit{Sparse connectivity in $G$}: problems with
 variable iteractions not matching the structure of $G$ cannot be solved directly.
\item \textit{Small numbers of total qubits $|V(G)|$}: only
  problems of up to 1100 variables can currently be addressed.
\end{enumerate}
\cite{Bian14} suggested approaches ameliorating these concerns. The
core idea used to address concerns 1 to 3 is the construction of
penalty representations of constraints with large (classical) energy gaps between
feasible and infeasible configurations. The large energy gaps buffer
against parameter error and maximize energy scales relative to the
fixed device temperature. Sparse device connectivity was addressed using \textit{locally-structured embedding}, which consists of placing constraints directly onto disjoint subgraphs of $G$ and routing constraints together using chains of ferromagnetically-coupled qubits representing the same logical variable. This differs from the more common global approach in which constraints are modelled without regard for local hardware structure. We contrast the two approaches in \S{\ref{subsec:approaches}}, and provide some experimental evidence that the locally-structured approach is well-suited to current QA hardware.

With locally-structured embedding, the number of qubits used, size of the energy gaps, and size of chains play an important role in determining D-Wave hardware performance. Here, we expand on the methods in \cite{Bian14} and offer several improvements. One way of reducing the required number of qubits, described in \S{\ref{sec:clustering}}, is by clustering constraints, thereby reducing the number of literals in the CSP. To maximize energy gaps, we follow the methods in \cite{Bian14} but extend them to max-constraint-satisfaction problems (MAX-CSP): given a set of constraints, find a variable assignment that minimizes the number of constraints that are unsatisfied. \S{\ref{sec:constraints}} describes two extensions: one that involves the explicit introduction of variables to indicate the reification of the constraints, and one that does not. Lastly, \S{\ref{sec:embedding}} describes how to reduce the size of the largest chains used by combining placement and routing into a single, iterative algorithm. Using linear programming, we can also find effective lower bounds on the size of the largest chains, which makes optimal routing faster.

To address the issue of a limited number of total qubits, \cite{Bian14} adapted two problem decomposition methods to the Ising context, namely dual decomposition (DD) and belief propagation (BP). However these algorithms suffer from issues of precision and a large number of iterations respectively. In \S{\ref{sec:decomposition}} we give two alternatives. One is the well-studied Divide-and-Concur algorithm \cite{Gravel08}, which produces excellent experimental results. The other is a novel message passing algorithm based on distributed minimization of the Bethe free energy called \textit{Regional Generalized Belief Propagation}; this offers some of the potential benefits of BP with far fewer calls to the QA hardware.

A salient feature of D-Wave QA device is the low cost of
 sampling low energy configurations of Eq.~\eqref{eq:Ising}. After a constant overhead time to program $\vec{h}$ and $\vec{J}$, additional i.i.d. samples can be obtained at an annealing rate of 20 $\mu$s/sample.  Consequently, problems where a diversity of ground states are sought form an interesting application domain. As an application of our MAX-CSP modelling techniques, we focus on the fault diagnosis problem. In fault diagnosis each constraint is realized as a logical gate which defines the input/output pairs allowed by the gate. A circuit of gates then maps global inputs to global outputs. An error model is prescribed for each gate, and fault diagnosis seeks the identification of a minimum number of faulty gates consistent with observed global inputs and outputs and error model. A diversity of minimal cost solutions is valuable in pinpointing the origin of the faults. In \S{\ref{sec:faults}}, we test the ability of the D-Wave hardware to generate a range of minimal cost solutions and also use the hardware to test various decomposition algorithms on a standard benchmark set of fault diagnosis problems. 

\section{Methods}\label{sec:methods}

\subsection{Approaches to Embedding}\label{subsec:approaches}

\begin{figure}[ht]
\centering
\includegraphics[scale=0.9]{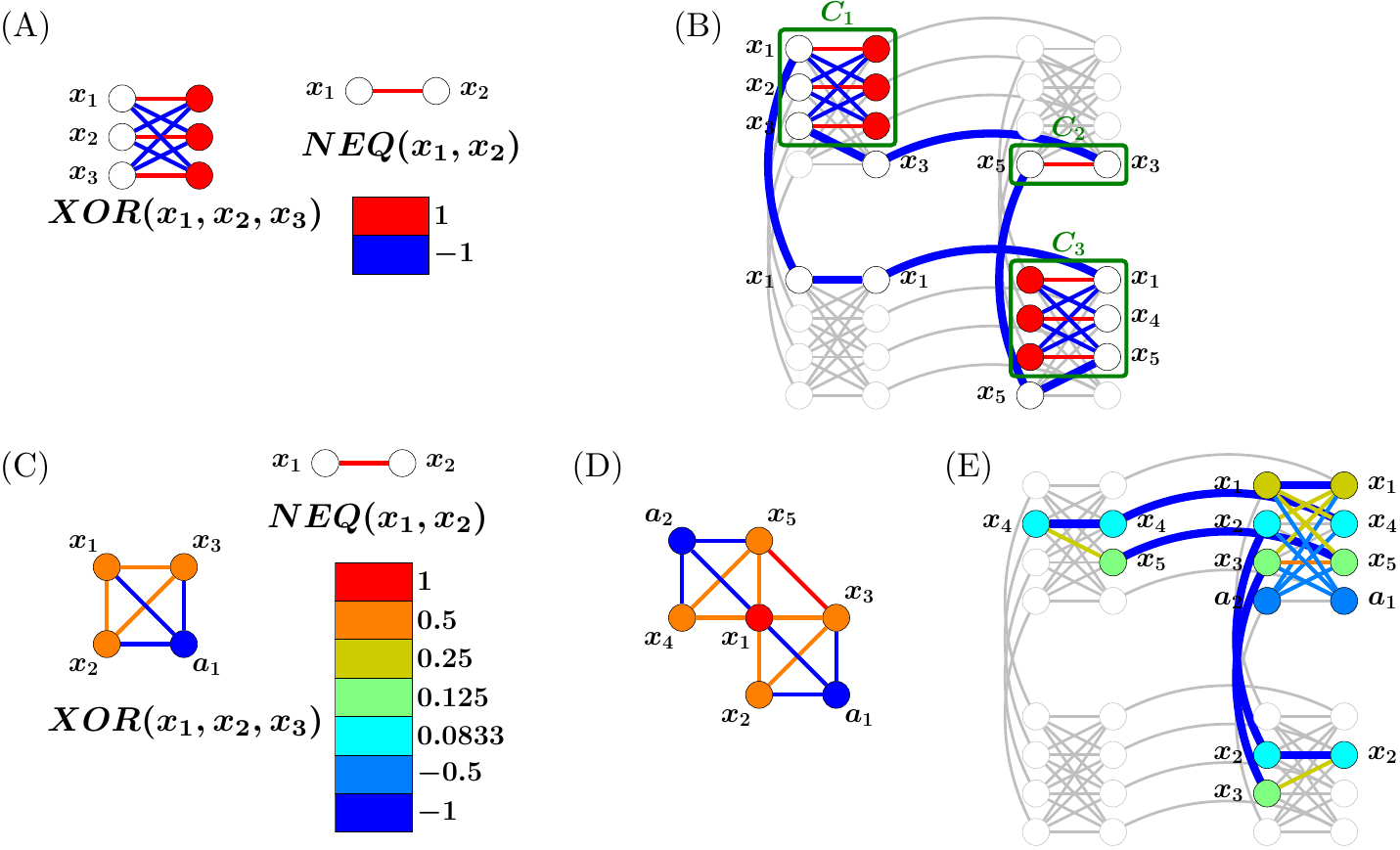} 
\caption{Comparison of locally-structured (top) and global (bottom) embeddings of a CSP with constraints $\{XOR(x_1,x_2,x_3), XOR(x_1,x_4,x_5),NEQ(x_3,x_5)\}$ in a D-Wave-structured hardware graph. (A) The penalty models for $XOR$ and $NEQ$ have an energy gap $g = 2$. (B) After locally-structured embedding with a chain strength of $\alpha = 1$, the Ising model for the CSP has an energy gap of $2$. Chain couplings are indicated with thick blue edges. (C) The given penalty model for $XOR$ using only 4 qubits has energy gap $g = 1$. (D) The aggregate Ising model for the CSP. Variable $a_i$ is an auxiliary variable used to define the $i$-th constraint. (E) Global embedding of the aggregate model. The chain strength $\alpha = 2$ was optimized experimentally, and the entire Ising model is scaled by a factor of $1/\alpha$ to satisfy the range requirement $-1\le J_{ij} \le 1$. After scaling, the Ising model for the CSP has an energy gap of $g = 0.5$. The global embedding uses fewer qubits but requires more precision to specify.}
\label{approachesFig}
\end{figure}

Modeling a constrained problem as a $G$-structured Ising objective
 requires reconciliation of the problem's structure with that of $G$. Two approaches may be taken to accommodate the connectivity required by $G$. 
 In \textit{global embedding} we model each constraint as an Ising model without regard for the connectivity of $G$, add all constraint models, and map the structure of the aggregate model onto $G$ using the heuristic minor-embedding algorithm of \cite{Cai14}.  Previous examples of global embedding include \cite{Bian13,Douglass15,Perdomo15,Rieffel15,Venturelli15,Zick15}. 
 Alternatively, when the scopes of constraints are small, \textit{{locally-structured embedding}} \cite{Bian14} models each constraint locally within a subgraph $\mathcal{G}\subset G$, places the local subgraphs $\mathcal{G}$ within $G$, and then connects (routes) variables occurring in multiple local subgraphs. Figure~\ref{approachesFig} contrasts the two approaches. 

 The methods offer different trade-offs. The former method typically
 utilizes fewer qubits, and has shorter \textit{chains} of connected qubits representing logical problem variables. The latter method is more scalable to large problems, usually requires less precision on parameters, and offers lower coupling strengths used to enforce chains. More precisely, assume an embedded Ising model is parameterized by $[\vec{h}, \vec{J}] = [\vec{h}, \vec{J}_P + \alpha \vec{J}_C]$, where $[\vec{h}, \vec{J}_P]$ describes the encoded constraints, $\vec{J}_C$ enforces the couplings within chains, and $\alpha > 0$ is a chain strength. In a satisfiable CSP that has been embedded with the locally-structured approach, the chains representing a solution to the CSP will be a ground state of $[\vec{h}, \vec{J}]$ regardless of the choice of $\alpha$.\footnote{To see this, note that $[\vec{h}, \vec{J}_P]$ is a collection of penalty models for constraints on independent variables, each of which achieves its ground state energy when the constraint is satisfied, while $[0, \vec{J}_C]$ achieves its ground state energy whenever no chains are broken.}  In contrast, for some global embeddings, the chain strength required to enforce unbroken chains can grow with system size, increasing the precision with which the original problem must be represented and making the dynamics of quantum annealing more difficult \cite{Venturelli15b}.

\begin{figure}[ht]
\centering
\includegraphics[scale=0.31]{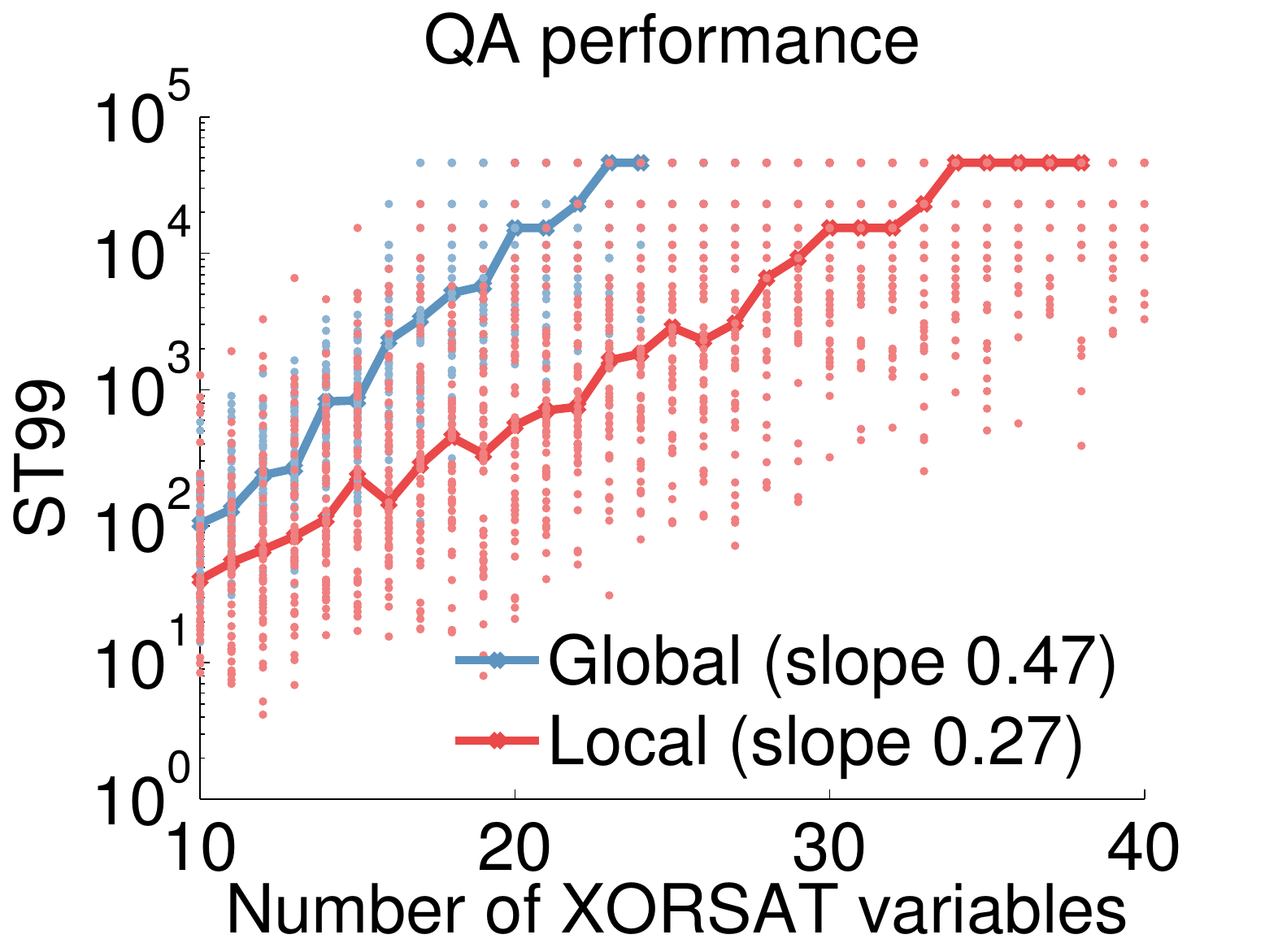}
\includegraphics[scale=0.31]{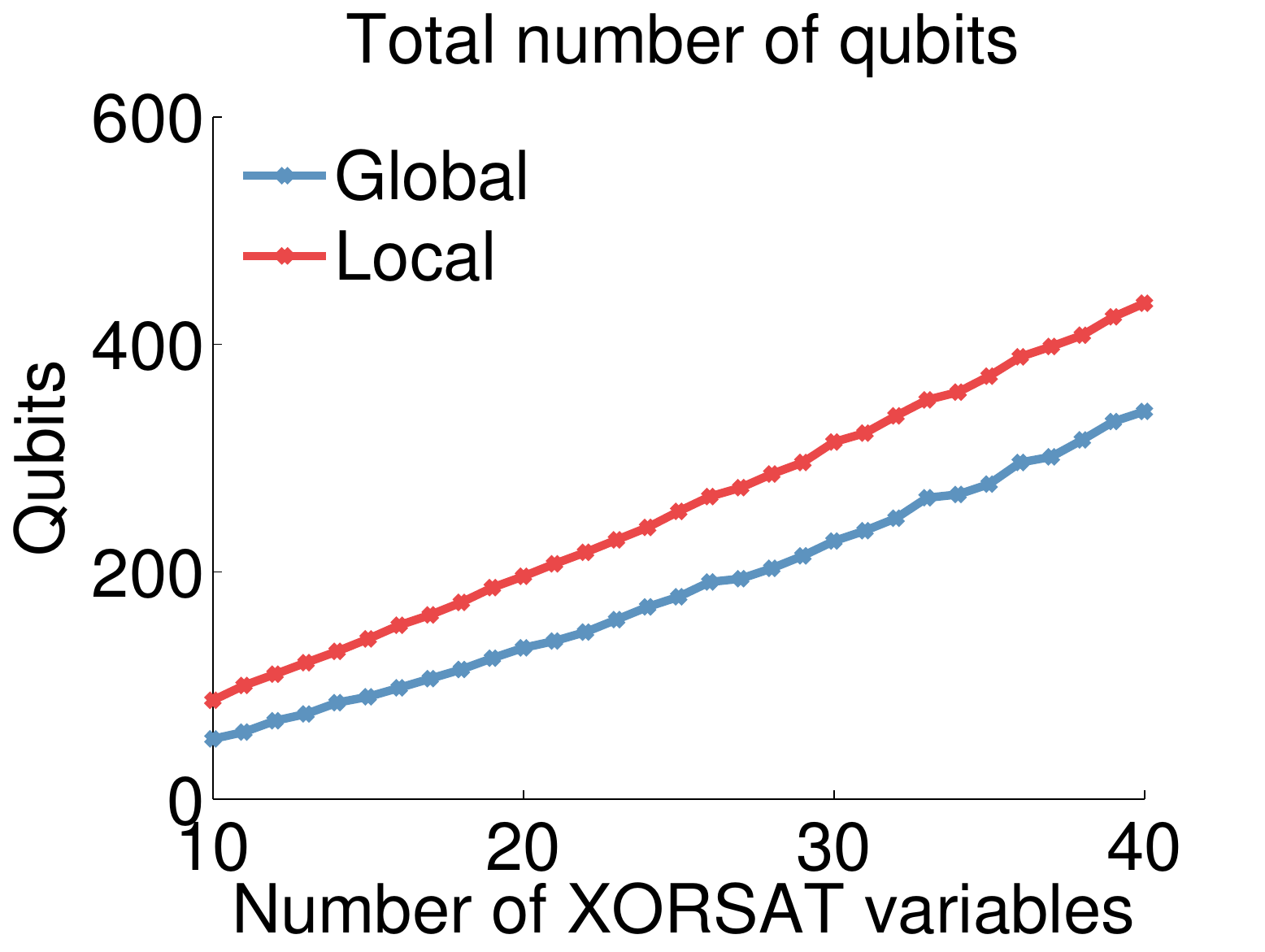}
\includegraphics[scale=0.31]{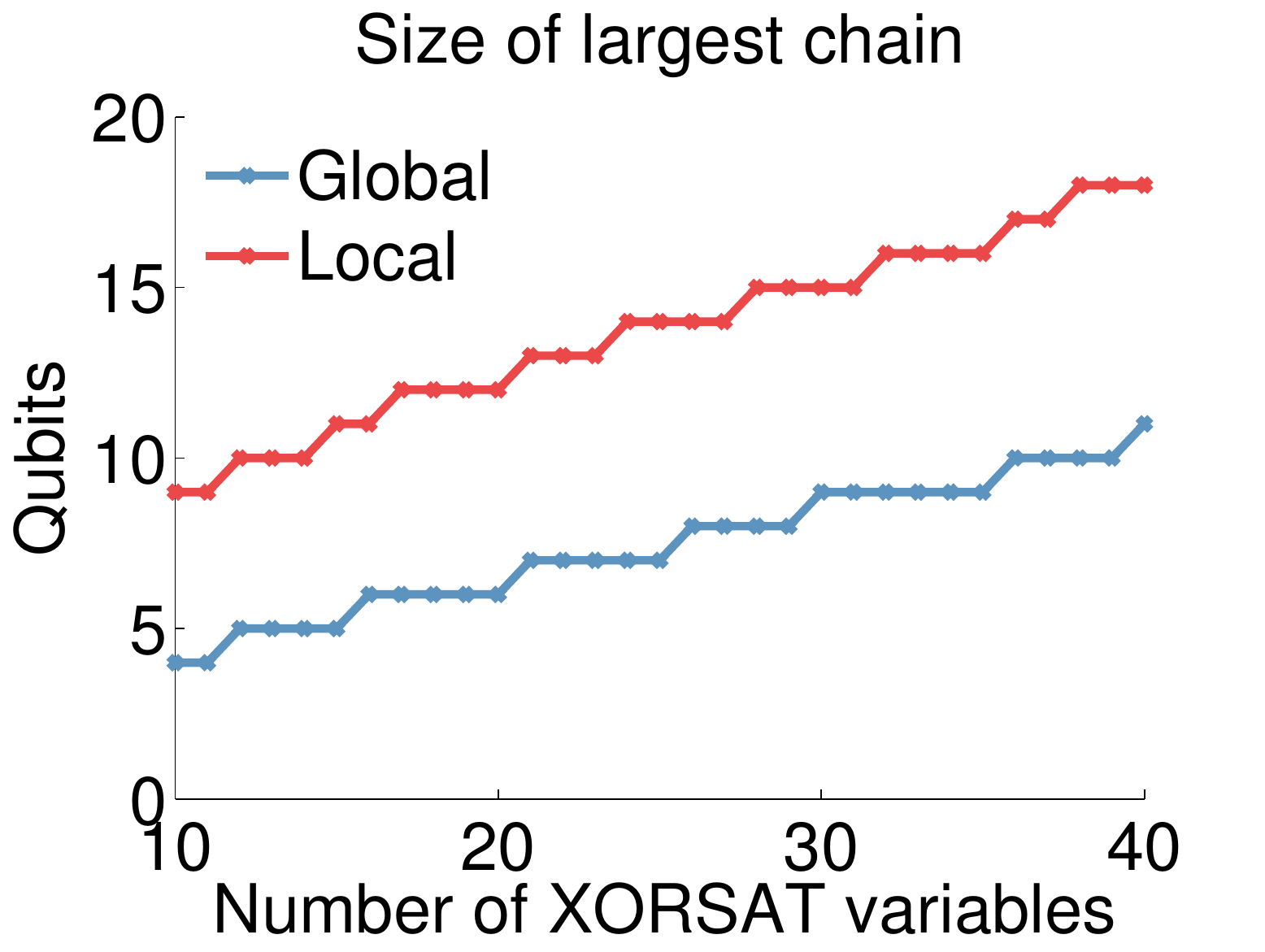}
\caption{Comparison of the D-Wave quantum annealing hardware performance in solving XOR-3-SAT problems \cite{Jia05}, using global (blue) and locally-structured (red) modelling strategies. Each XOR-3-SAT instance is randomly generated subject to having a unique solution and a clause-to-variable ratio of 1.0. Global embeddings: each constraint $s_1s_2s_3 = 1$ (with $s_i = \pm 1$) is encoded in the 4-qubit penalty model in Figure~\ref{approachesFig}(C), which has energy gap $g = 1$. The Ising model representing the sum of these constraints is then embedded using the heuristic in \cite{Cai14}. Local embeddings: each constraint is mapped directly to the $K_{3,3}$-structured Ising model in Figure~\ref{approachesFig}(A), which has energy gap $g = 2$. Constraints are then embedded using the rip-up and replace algorithm in \S{\ref{subsubsec:rrr}}. 
Left: Scaling of hardware performance with no post-processing. ST99 is the number of samples needed to find the ground state with 99\% probability, which, given a fraction $p$ of all samples taken that are in the ground state, is given by the formula ST99 = $\log(1-.99)/\log(1-p)$ \cite{Ronnow14}. Bold lines indicate the median across $50$ instances of each problem size. For points not shown, the hardware failed to find a ground state. Middle: total number of qubits required to embed the Ising model (median). Right: number of qubits in the largest chain in the embedding (median).}
\label{fig:XORHW}
\end{figure}

Whether the benefits of improved precision and lower chain strengths outweighs the drawbacks of using more qubits depends on the problem. Figure~\ref{fig:XORHW} gives an example of a problem class (random XOR-3-SAT problems) for which the overall performance and scaling of quantum annealing hardware is noticeably improved with locally-structured embedding. For the fault diagnosis problems studied here, the locally-structured approach also performs better, and we pursue improvements to the locally-structured algorithm of \cite{Bian14}.

\subsection{Preprocessing}\label{sec:clustering}

For some CSPs, aggregating several small constraints into a single
 larger one prior to modelling may lead to more efficient hardware mappings and better hardware performance. The benefit stems from the fact that variables appearing in multiple constraints within a cluster need only be represented once (or perhaps not at all). As an example, consider the boolean-valued constraint $y = \text{XOR}(x_1,x_2)$. If XOR is represented using AND/OR/NOT gates, for example as $\{a_1 = \text{AND}(x_1,\neg x_2), a_2 = \text{AND}(x_2,\neg x_1), y = \text{OR}(a_1,a_2)\}$, at least $9$ literals are needed. On the other hand, by clustering the three gates XOR can be represented by an Ising model directly using only $4$ qubits (see Figure~\ref{approachesFig}(c)).

Unfortunately as constraints become larger it becomes more difficult to find Ising models to represent
them. This upper limit on the practical cluster size requires straightforward modifications to standard clustering methods like agglomerative clustering \cite{Tan05}. When the CSP is derived from a combinational circuit, cone-based clustering \cite{Siddiqi07,Metodi14} can also be adapted to accommodate bounds on the cluster size. Both agglomerative and cone clustering can be performed in polynomial time.

\subsection{Mapping constraints to Ising models}
\label{sec:constraints}

Regardless of whether or not constraints are clustered, the next step in mapping to hardware is identifying an Ising model to represent each constraint. We assume a constraint on $n$ binary variables is characterized by a subset of $\{0,1\}^n$ which indicates valid variable assignments.  Since quantum hardware uses spin variables with values in $\{-1,+1\}$, we identify 0 with -1 and 1 with +1, and assume that the feasible set $\vec{F}$ is a subset of $\{-1,+1\}^{n}$.  Our goal is to find an Ising model that separates the feasible solutions $\vec{F}$ from the infeasible solutions $\{-1,+1\}^{n} \setminus \vec{F}$ based on their energy values. In particular, the ground states of the Ising model must coincide with the feasible solutions $\vec{F}$. Furthermore, to improve hardware performance, we seek Ising models for which the gap, that is, the smallest difference in energy between feasible and infeasible solutions, is largest.

Typically, due to both the complexity of the constraint and the
sparsity of the hardware graph, the Ising model requires ancillary
variables, which also may help to obtain larger gaps.  We assume
that the allowable interactions in the Ising model are given by an $m$-vertex subgraph $\mathcal{G}$ of the hardware graph $G$, where $m\ge n$. The
constraint variables are mapped to a subset of $n$ vertices of $\mathcal{G}$, while the rest
of the vertices are associated with $h=m-n$ ancillary variables. For simplicity, we write a spin configuration $\vec{z} \in
\{-1,+1\}^{m}$ as $\vec{z}=(\vec{s},\vec{a})$, meaning that the working
variables take the values $\vec{s}$, while the ancillary variables are
set to $\vec{a}$.

The Ising model we seek is given by variables $\vec{\theta} = (
\theta_0, (\theta_i)_{i\in V(\mathcal{G})}, (\theta_{i,j})_{(i,j)\in
  E(\mathcal{G})})$, where $\theta_i$ are the local fields $h_i$,
$\theta_{i,j}$ are the couplings $J_{i,j}$, and $\theta_0$ represents
a constant energy offset (unconstrained).  To simplify the notation,
for a configuration $\vec{z}$ we define $\vec{\phi}(\vec{z}) = ( 1,
(z_i)_{i\in V(\mathcal{G})}, (z_i z_j)_{(i,j)\in
  E(\mathcal{G})})$. Thus, the energy of $\vec{z}$ is given by
\[
 \mathbb{E}_{\vec{\theta}}(\vec{z}) = \ip{\vec{\theta}}{\vec{\phi}(\vec{z})}.
\]
The hardware imposes lower and upper bounds on $\vec{\theta}$, that we
denote respectively by $\underline{\vec{\theta}}$ and
$\overline{\vec{\theta}}$ (with $\overline{\theta}_0 = +\infty$
and $\underline{\theta}_0=-\infty$).  Current D-Wave hardware
requires $h_i\in [-2,2]$ and $ J_{i,j}\in[-1,1]$.

To separate feasible and infeasible solutions we require that for some positive gap $g$:
\begin{equation}
  \min_{\vec{a}} \mathbb{E}_{\vec{\theta}}(\vec{s},\vec{a}) = 0 \ ,  \quad \forall \vec{s}\in \vec{F}\qquad  \text{and} \qquad
  \min_{\vec{a}} \mathbb{E}_{\vec{\theta}}(\vec{s},\vec{a}) \ge g \ , \quad  \forall \vec{s}\not\in \vec{F}. 
\end{equation}

Thus, the problem of finding the Ising model with largest gap can be stated as follows
\begin{align}
  \max_{g,\vec{\theta}} & \;\; g \notag \\
  \text{subject to} \;\;\qquad
  \ip{\vec{\theta}}{\vec{\phi}(\vec{s},\vec{a})} &\ge 0 && \forall
  \vec{s}\in
  \vec{F}, \forall \vec{a} \label{cons1} \\
  \;\; \ip{\vec{\theta}}{\vec{\phi}(\vec{s},\vec{a})} &\ge g && \forall
  \vec{s}
  \not\in \vec{F}, \forall \vec{a} \label{cons2} \\
  \;\; \exists \ \vec{a} : \
  \;\ip{\vec{\theta}}{\vec{\phi}(\vec{s},\vec{a})} &= 0 && \forall
  \vec{s}\in \vec{F} \label{cons3Eq} \\
  \;\; \underline{\vec{\theta}} \le &\vec{\theta} \le
  \overline{\vec{\theta}}. &&\notag
\end{align}
Here, constraints \eqref{cons1} and \eqref{cons3Eq} guarantee that all
feasible solutions have minimum energy 0, while constraint
\eqref{cons2} forces infeasible solutions to have energy at least $g$.

This optimization problem is solved as a sequence of feasibility
problems with fixed gaps $g$.  Using the fact that the interaction graph $\mathcal{G}$ has low tree-width, we
can significantly condense the formulation above. In this way the
number of constraints may be reduced from exponential in $m$ to
exponential in the treewidth of $\mathcal{G}$. The resulting model is
solved with a Satisfiability Modulo Theories (SMT) solver (see
\cite{Bian14} for more details).

The penalty-finding techniques above assume that a placement of
variables within the Ising model is given. However different
placements allow for different energy gaps, and it is not clear, even
heuristically, what characteristics of a placement lead to larger
gaps. For small constraints, canonical augmentation \cite{Mckay98} can be used to generate all non-isomorphic placements. 

\subsubsection{Methods for MAX-CSP}
\label{sec:max-csp}

Borrowing from fault diagnosis terminology,
we consider constraints characterized by two disjoint subsets of feasible
solutions: \emph{healthy} states $\vec{F}_1 \subseteq \{-1,1\}^{n}$, and \emph{faulty} states $\vec{F}_2 \subseteq \{-1,1\}^{n}$. States in $\{-1,1\}^{n} \setminus ( \vec{F}_1 \cup \vec{F}_2 )$ are considered infeasible.  As before, we require an Ising
model that separates feasible from infeasible solutions, but preferring
healthy to faulty states whenever possible. The particular case
$\vec{F}_2 = \{-1,1\}^{n} \setminus \vec{F}_1$ corresponds to a MAX-CSP
problem, in which a CSP is unsolvable but nonetheless we attempt to
maximize the number of constraints satisfied by applying the same penalty
to every constraint with a faulty configuration.

One way to model $(\vec{F}_1,\vec{F}_2)$ is through reification where a variable representing the truth of the constraint is introduced. This reified, or health, variable is +1 for
healthy states and -1 for faulty states. That is, we define a feasible set
\[
\vec{F} = \{ (\vec{x},+1): \vec{x} \in \vec{F}_1 \} \cup \{ (\vec{x},-1):
\vec{x} \in \vec{F}_2 \} \subseteq \{-1,1\}^{n+1} ,
\]
and model $\vec{F}$ using the methods in \S{\ref{sec:constraints}}.
In this case, both solutions in $\vec{F}_1$ and $\vec{F}_2$ will be
equally preferred. To break the tie to favor healthy states, the
health variable can be added to the objective function with negative
weight.\footnote{More generally, weighted CSP can be solved by weighting reified variables representing constraints.}  We call this the \emph{explicit} fault model.

A second strategy is to modify the optimization problem of \S{\ref{sec:constraints}}
so that all solutions in $\vec{F}_1$ have energy 0, while all solutions
in $\vec{F}_2$ have energy $e>0$ and infeasible solutions have
energy at least $g>e$. In this case, we fix the intermediate energy
$e$ and the optimization problem becomes:
\begin{align}
  \max_{\vec{\theta}, g\ge e}  \;\; g & &&\notag \\ 
  \text{subject to}  \;\;\qquad
  \ip{\vec{\theta}}{\vec{\phi}(\vec{s},\vec{a})} &\ge 0  &&\forall \vec{s}\in
  \vec{F}_1, \forall \vec{a} \label{cons1-2} \\
   \;\;
  \ip{\vec{\theta}}{\vec{\phi}(\vec{s},\vec{a})} &\ge e &&\forall \vec{s}
  \in \vec{F}_2, \forall \vec{a} \label{cons12-2} \\ 
  \;\;
  \ip{\vec{\theta}}{\vec{\phi}(\vec{s},\vec{a})} &\ge g  &&\forall \vec{s}
  \not\in \vec{F}_1 \cup \vec{F}_2, \forall \vec{a} \label{cons2-2} \\ 
   \;\; \exists \ \vec{a} : \
  \;\ip{\vec{\theta}}{\vec{\phi}(\vec{s},\vec{a})} &= 0  &&\forall
  \vec{s}\in \vec{F}_1 \label{cons3Eq-2} \\ 
   \;\; \exists \ \vec{a} : \
  \;\ip{\vec{\theta}}{\vec{\phi}(\vec{s},\vec{a})}& = e  &&\forall
  \vec{s}\in \vec{F}_2 \label{cons3Eq-12} \\ 
  \;\; \underline{\vec{\theta}}
  \le &\vec{\theta} \le \overline{\vec{\theta}}. \notag 
\end{align}
It is straightforward to adapt the SMT solution methods of
\cite{Bian14} to this problem. We call this the \emph{implicit} fault model. The implicit model generally requires fewer variables (i.e.,
qubits). However, care must be taken to ensure that $g$ is large compared to $e$; otherwise, when adding penalties together, it may be difficult to differentiate several faulty constraints from a single infeasible constraint. In the explicit model, this issue can be avoided  by choosing a sufficiently small weight for health variables in the objective function.

\subsection{Locally-structured embedding}
\label{sec:embedding}

Given a method for generating penalties on subgraphs $\mathcal{G}$ the next steps of locally-structured embedding are the placement of $\mathcal{G}$'s within $G$, and the routing of chains of interactions between variables occurring in multiple constraints.  \cite{Bian14}
suggested adapting VLSI algorithms for placement \cite{Kahng11,Chan00,Roy06} and routing \cite{Kahng11,Gester13} to accomplish these steps, and in this section, we describe two improvements to that work. Firstly, using routing models we find a tight lower bound on the size of the largest chain. This bound is combined with search heuristics to speed the discovery of good embeddings.  Secondly, embedding algorithms that utilize placement and routing steps differ in a significant way from their classical VLSI counterparts, and a modification that performs simultaneous placement and routing improves results.

\subsubsection{Chain length lower bounds and improved routing}
\label{lowerbounds}

The performance of D-Wave's hardware in solving an embedded Ising
 model depends heavily on the size of the chains of variables: shorter chains are more likely to yield better performance \cite{Venturelli15b}. In this section, we focus on routing which minimizes chain lengths. We assume that constraints have already been placed in the hardware (see \cite{Bian14} for placement methods).  We show how to find tight lower bounds on the maximum chain size in an embedding, and provide an effective procedure to improve routing using these bounds.

We first consider bounds for a single chain, which reduces to the well-studied Steiner tree problem.
Let $T\subseteq V(G)$ be a set of \emph{terminals}; i.e. qubits in the hardware graph to which a variable has been assigned during placement. A \emph{Steiner tree} is a connected subgraph of $G$ that contains all the terminals. The Steiner tree problem consists of finding the smallest (fewest number of nodes) Steiner tree. The non-terminal vertices in a Steiner tree are called Steiner points.

There are several ways to model the Steiner tree problem as an mixed
 integer linear program (MILP). However the tightness of the linear program (LP) relaxation will have a significant impact on the time required to find a solution. Here we consider a formulation whose LP relaxation, known as the \emph{bidirected cut relaxation}, has an integrality gap of at most 2 \cite{Rajagopalan99}.  First, we transform $G$ into a directed graph by replacing each edge with two opposite arcs. For each $v \in V(G)\setminus T$, let $x_v$ be a binary variable indicating whether $v$ is part of the Steiner tree. When variables $x_v$ are fixed, the Steiner tree is just a tree spanning $T \cup \{v\in V(G) : x_v=1\}$.  A tree can be modelled as a multi-commodity transshipment problem: pick any $v_0\in T$ as root, and find a path from $v_0$ to each of the other $|T|-1$ terminals.  Concretely, if we define flow variables $f_a^i$ indicating that arc $a$ is on the path from $v_0$ to terminal $i$, then an MILP formulation for the the Steiner tree problem is
\begin{align}
 \text{(BCR)} \qquad \qquad \min & \;\; \sum_{v\in V\setminus T} x_v \notag \\ \text{subject to} & \;\;
 \sum_{v\rightarrow a} f_a^i -  \sum_{a\rightarrow v} f_a^i = \begin{cases}
                                                       1 &\text{if $v=v_0$}\label{trans-flow2}\\
-1&\text{if $v=v_i\in T, i\not=0$}\\
0  &\text{if $v\notin T$}  
                                                  \end{cases} \\
& \;\; \sum_{a\rightarrow v} f_a^i \le  x_v & \forall {v\in V\setminus T}, i\not=0 \label{cap2}\\
 & \;\;
  0 \le f_a^i \le 1 & \forall a, i\not=0 \notag \\ 
& \;\;  x_v \in \{0,1\} & \forall {v\in V\setminus T} \notag 
 \notag 
\end{align}
Here, constraints \eqref{trans-flow2} are the flow constraints, while the capacity constraints \eqref{cap2} allow flow to be routed only through Steiner points (i.e., $v\in V(G)$ with $x_v=1$). The notation ${v\rightarrow a}$ (respectively ${a\rightarrow v}$) refers to all arcs whose tail is $v$ (respectively, whose head is $v$).

The LP relaxation of program (BCR) above produces very tight lower bounds for a range of Steiner tree problems \cite{Chopra92}. This MILP can be extended to the full routing problem using different flows for each Steiner tree to be found, with the additional demand that every variable can appear in at most one Steiner tree. 

Having access to good bounds on the chain lengths allows for a simple improvement to the routing phase presented in \cite{Bian14}. Assume we have a heuristic routing algorithm {\sc Route$(G,\mathcal{T},M)$} that takes as input a hardware graph $G$, a collection of terminal sets $\mathcal{T} = \{T_i\}$ for each variable $x_i$ in the CSP, and a maximal allowable chain size $M$. Then any successful call to {\sc Route$(G,\mathcal{T},M)$} will provide an upper bound on the maximal chain length no worse than $M$, while any unsuccessful call provides a lower bound of $M+1$. With these bounds we can perform a heuristic binary search for the optimal maximal chain length, and beginning with the good lower bound provided by the LP relaxation of (BCR) will significantly reduce the number of iterations in the search.



\subsubsection{Combined place-and-route algorithms}\label{subsubsec:rrr}

The place-and-route model of embedding, while known to scale well, is
 often inefficient in maximizing the size of a problem embeddable in a fixed hardware graph. One reason is that in contrast with VLSI, the resources being negotiated by placement and routing are identical (namely, vertices of $G$). So, for example, an optimal placement might squeeze constraints as close together as possible, leaving no room for routing.  For this reason we have developed a \textit{rip-up-and-replace} algorithm which combines the placement and routing phases of embedding, using new routing information to update placements and vice versa.

During the course of the algorithm, vertices of $G$ may be temporarily assigned to multiple variables, and each vertex is given an exponentially increasing penalty weight according to the number of times it has been used. More precisely, if each variable $x_i$ currently has chain $S_i \subset V(G)$, then the weight of vertex $q \in V(G)$ is $\omega(q) = \alpha^{|\{i: q \in S_i\}|}$ for some fixed $\alpha > 1$. Each constraint $C$ is given a placement $(L_C,v_C)$ consisting of a location $L_C \subset V(G)$ and an assignment of variables to vertices within the location, $v_C: V(C) \rightarrow L_C$ (where $V(C)$ denotes the set of variables associated with constraint $C$).

The algorithm iteratively alternates between assigning constraints to locations, and routing variables between constraints (\textit{i.e.} creating chains). Chains are created using a weighted Steiner tree approximation algorithm such as the MST algorithm \cite{Kou81} or Path Composition \cite{Gester13}. Constraint locations are chosen based on a cost function, where the cost of $(L,v)$ depends on the weight of vertices in $L$ and the weight of routing to $(L,v)$, which is approximated by weighted shortest-path distances to existing chains. The algorithm terminates when a valid embedding is found or no improvement can be made. Explicit details are given in Algorithm \ref{fig:rrralg} below.

\begin{algorithm}
\begin{algorithmic}
\Require{Graph $G$, list of constraints $\mathcal{C}$, list of potential placements $(L,v)$ for each $C \in \mathcal{C}$}
\Ensure{Placement of each constraint $C \in \mathcal{C}$ on a location $(L_C,v_C)$ and a chain $S_i \subset V(G)$ for each variable $x_i$ such that all chains are disjoint, or ``failure".} 
\Statex
\Function{RipUpAndReplace}{$G$,$\mathcal{C}$}
\State Choose an initial placement $(L_C,v_C)$ for each $C \in \mathcal{C}$
\For{each variable $x_i$}
	\State{$T_i \gets \{v_C(x_i): C\in \mathcal{C}, x_i \in V(C)\}$}
	\State{$S_i \gets$ approximately minimal Steiner tree for terminals $T_i$}
	\EndFor
\For{$q \in V(G)$}
	\State $\omega(q) \gets \alpha^{|\{i: q \in S_i\}|} \quad$ 	 (for fixed $\alpha > 1$)
	\EndFor
\While{$\max_{q \in V(G)} |\{i: q \in S_i\}|$ is improving}
	\State Randomize the order of $\mathcal{C}$
	\For{each $C \in \mathcal{C}$}
		\For{$x_i \in V(C)$}
			\State{$S_i \gets $ \Call{Trim}{$S_i,C$}}
			\State{Update $\omega(q)$ for $q \in S_i$}
			\State{Compute $d(q,S_i) \gets \omega$-weighted shortest-path distance from $S_i$ to $q$, $\forall q \in V(G)$}
			\EndFor
		\For{each potential location $(L,v)$ for $C$}
			\State{$\cost(L,v) \gets \sum_{q \in L} \omega(q) + \sum_{x_i \in C} d(v(x_i),S_i)$}
			\EndFor
		\State{Pick new location $(L_C,v_C) \gets (L,v)$ for $C$ with probability $\propto \beta^{-cost(L,v)}$ (fixed $\beta > 1$)}
		\State{Update $\omega(q)$ for $q \in L_C$}
		\For{$x_i \in V(C)$}
			\State{$T_i \gets \{v_{C'}(x_i): C' \in \mathcal{C}, x_i \in V(C')\}$}
			\State{$S_i \gets \omega$-weighted approximately minimal Steiner tree for $T_i$}
			\State{Update $\omega(q)$ for $q \in S_i$}
			\EndFor
		\EndFor
	\EndWhile
	\If{$\max_{q \in V(G)} |\{i: q \in S_i\}| = 1$}
		\State{Optimize chain length of chains $\{S_i\}$ for terminals $\{T_i\}$}
	\Else
		\State{Return ``failure"}
		\EndIf
\EndFunction		
\Function{Trim}{$S_i$,$C$}
	\State{$T_i \gets \{v_{C'}(x_i): C' \neq C, x_i \in V(C')\}$}
	\While{some $x \in S_i \backslash T_i$ has degree $1$ in the subgraph of $G$ induced by $S_i$}
		\State{$S_i \gets S_i \backslash \{x\}$}
	\EndWhile
	\State{Return $S_i$}
\EndFunction
\end{algorithmic}
\caption{Rip-up and replace heuristic for finding a placement of constraints and embedding of variables in a hardware graph.}
\label{fig:rrralg}
\end{algorithm}

Alternatives to rip-up-and-replace, which iteratively updates the locations for constraints based on variable routing, such as simulated annealing or genetic algorithms could be used to update placements. For example, define a gene to consist of a preferred location for each constraint, and a priority order for constraints. Given a gene, constraints are placed in order of priority, in their preferred location if it is available or the nearest available location otherwise. During simulated annealing, genes are mutated by perturbing the preferred location for a constraint or transposing two elements in the priority order. These algorithms tend to take much longer than rip-up-and-replace, but eventually produce very good placements. 

\subsection{Decomposition algorithms}
\label{sec:decomposition}

Owing to a limited number of qubits, it is often the case that a CSP or Ising model is too large to be mapped directly onto the hardware. \cite{Bian14} offered various decomposition techniques which use QA hardware to solve subproblems as a subroutine for solving larger ones. In this section, we describe two additional algorithms: \textit{divide-and-concur} \cite{Gravel08,Yedidia11}, specialized to our case of Ising model energy minimization, and a new algorithm inspired by regional generalized belief propagation \cite{Yedidia2005}.

For both algorithms, we partition the constraints of a MAX-CSP into regions $\mathcal{R}=\{R_1,R_2,\cdots\}$ so that each subset of constraints can be mapped to a penalty model on the hardware using the methods of the previous section. For a region $R \in \mathcal{R}$, the penalty model $[\vec{h}^{(R)},\vec{J}^{(R)}]$ produces an Ising energy function $E_R(\vec{z}^{(R)})$ whose ground states satisfy all the constraints in that region. Here $\vec{z}^{(R)}$ is the subset of variables involved in the constraints of region $R$. Since embedding is slow in general, regions are fixed and embedded in hardware as a pre-processing step.

The key problem is that sampling a random ground state from each region produces inconsistent settings for variables involved in multiple regions' constraints. At a high level, messages passed between regions indicate beliefs about the best assignments for variables, and these are used to iteratively update the biases on $\vec{h}^{(R)}$ in hopes of converging upon consistent variable assignments across regions. The two algorithms presented here implement this strategy in very different ways.

\subsubsection{Divide and concur (DC)}

Divide-and-concur \cite{Gravel08,Yedidia11} (DC) is a simple message passing algorithm that attempts to resolve discrepancies between instances of variables in different regions via averaging. In each region $R$, in addition to having an an Ising model energy function $E_R(\vec{z}^{(R)})$ representing its constraints, one introduces linear biases $L_R(\vec{z}^{(R)})$ on its variables, initially set to $0$. Let $z_i^{(R)}$ denote the instance of variable $z_i$ in region $R$. The two phases of each DC iteration are:
\begin{itemize}
\item Divide: minimize $E_R(\vec{z}^{(R)}) + L_R(\vec{z}^{(R)})$ in each $R$ (\textit{i.e.} satisfy all constraints and optimize over linear biases).
\item Concur: average the instances of each variable: $\overline{z}_i = \avg_{R: z_i \in R} z_i^{(R)}$ and update the linear biases by setting $L_R(\vec{z}^{(R)}) = \sum_{i \in R} -\overline{z}_i z_i^{(R)}$.
\end{itemize}
In the divide phase, $E_R$ is scaled appropriately so that the minimum of $E_R(\vec{z}^{(R)}) + L_R(\vec{z}^{(R)})$ satisfies all constraints.

This basic algorithm tends to get stuck cycling between the same states; one mechanism to prevent this problem is to extend DC with difference map dynamics \cite{Yedidia11}. DC has been shown to perform well on constraint satisfaction problems and constrained optimization problems, and compared to other decomposition algorithms, has relatively low precision requirements for quantum annealing hardware. That is, assuming each variable is contained in a small number of regions, the linear biases on the variables in the Ising model of each region (namely $-\overline{z}_i$) are discretized. On the other hand, like most decomposition algorithms, DC is not guaranteed to find a correct answer or even converge.

\subsubsection{Regional Generalized Belief Propagation (GBP)}

\cite{Bian14} explored min-sum belief propagation as a decomposition method. Here we take a different approach: instead of minimizing the energy of an Ising model $E(\vec{z})$ directly, we sample from its Boltzmann distribution $p(\vec{z}) = e^{-E(\vec{z})/T}/Z$. Presuming that we have successfully mapped our constraints to Ising models with large gaps (\S{\ref{sec:constraints}}), and that the temperature $T$ is sufficiently small, we have confidence that sampling from the Boltzmann distribution provides good solutions to the original constrained optimization problem. The Boltzmann distribution is the unique minimum of the Helmholtz free energy
$$A(p) = U(p) - TS(p) = \sum_{\vec{z}} p(\vec{z}) E(\vec{z}) + T \sum_{\vec{z}} p(\vec{z}) \log p(\vec{z}).$$
Our algorithm decomposes $A$ into regional free energies. The resultant algorithm is similar in spirit to the generalized belief propagation algorithm of \cite{Yedidia2005} based on their region graph method.

Sum-product belief propagation is related to critical points of the (non-convex) Bethe approximation, which for Ising energies reads
\begin{eqnarray*}
A_\mathrm{Bethe}(\{b_i\},\{b_{ij}\}) &=& \sum_{(i,j)\in E} \sum_{z_i,z_j=\pm 1} b_{ij}(z_i,z_j) J_{i,j}z_iz_j + T b_{ij}(z_i,z_j) \log b_{ij}(z_i,z_j)\\
&& \quad +\ \sum_{i\in V} \sum_{z_i=\pm 1} b_i(z_i)h_iz_i + T(1-d_i) b_i(z_i)\log b_i(z_i),
\end{eqnarray*}
where $d_i = |\{j\in V\::\: (i,j) \in E\}|$. The distribution $p$ in the free energy is approximated by local beliefs (marginals) $b_i,b_{ij}$ at each vertex and edge. To obtain consistent marginals, $b_i(z_i) = \sum_j b_{ij}(z_i,z_j)$ whenever $(i,j) \in E$, one introduces a constrained minimization problem, and it is the Lagrange multipliers associated to these constraints that relate to the fixed points of belief propagation. In particular, if belief propagation converges then we have produced an interior stationary point of the constrained Bethe approximate free energy~\cite{Yedidia2005}.

In our case, having divided a MAX-CSP into regions $\mathcal{R}$, we can formulate a regional analogue of the Bethe approximation,
\begin{eqnarray}\label{eqn:Bethe}
A^{\mathcal{R}}_\mathrm{Bethe}(\{b_i\},\{b_{R}\}) &=& \sum_{R\in\mathcal{R}} \left(\sum_{\vec{z}^{(R)}} b_R(\vec{z}^{(R)})E_R(\vec{z}^{(R)}) + T\sum_{\vec{z}^{(R)}} b_R(\vec{z}^{(R)})\log b_R(\vec{z}^{(R)})\right)\\\nonumber
&& \qquad + T \sum_{i} \left((1-c_i)\sum_{z_i} b_i(z_i)\log b_i(z_i)\right),
\end{eqnarray}
where now $c_i = |\{ R \::\: i\in R\}|$ is the number of regions whose Ising model includes variable $z_i$. In exactly the same way as above, requiring consistent marginals induces a constrained minimization problem for this regional approximation. The critical points of this problem are fixed points for a form of belief propagation. Specifically, for each variable $z_i$ in a constraint of $R$, the messages passed between variable and region are
\begin{eqnarray*}
m_{R\to i}(z_i) &\propto& \sum_{\vec{z}^{(R)}\setminus z_i} e^{-E_R(\vec{z}^{(R)})/kT} \prod_{j \in R \setminus i} m_{j\to R}(z_j)\\
m_{i\rightarrow R}(z_i) &\propto& \prod_{S \ni i \::\: S \not= R} m_{S\to i}(z_i)
\end{eqnarray*}
For large regions, which involve a large number of variables, the first of these messages is intractable to compute. As in previous work \cite{Bian14}, we use QA hardware to produce this message. In that work, the algorithm relied on minimizing the energy of the penalty model; here we harness the ability of the hardware to sample from the low energy configurations of the Ising model without relying on finding a ground state.

Unfortunately, it is not as simple as sampling from the Ising model formed from the constraints in a given region. Even if the hardware were sampling from its Boltzmann distribution, this would minimize the free energy of just that region
\begin{equation} \label{eq:regionalFreeEnergy}
A_R(p_R) = \sum_{\vec{z}^{(R)}} p_R(\vec{z}^{(R)})E_R(\vec{z}^{(R)}) + T\sum_{\vec{z}^{(R)}} p_R(\vec{z}^{(R)})\log p_R(\vec{z}^{(R)}).
\end{equation}
Unless the region $R$ is isolated, this would not recover the desired belief $b_R$ as we have failed to account for energy contributions of variables involved in other regions' constraints. We instead add corrective biases to each region's penalty model
\begin{equation}\label{eqn:Ising_correction}
\tilde{E}_R\left(\vec{z}^{(R)};\{V^{(R)}_j\}\right) = \sum_{(i,j)\in E^{(R)}} J^{(R)}_{ij} z_iz_j + \sum_{i\in V^{(R)}} h^{(R)}_i z_i + \sum_{i\in \partial R} V^{(R)}_i z_i
\end{equation}
and sample from the Boltzmann distribution of this energy function. We use the notation $E^{(R)}$ and $V^{(R)}$ for the Ising model graph associated to region $R$, and $\partial R \subset V^{(R)}$ for it's boundary: indices of variables that also appear in the penalty models of constraints in other regions. Only these variables gain corrective biases.

\begin{algorithm}
\begin{algorithmic}
\Require{A decomposition of a CSP into constraint regions $\mathcal{R}$ and penalty Ising models $E_R$ for each $R\in\mathcal{R}$. Putative temperature $T$.}
\Ensure{A critical point of the constrained regional Bethe approximation (\ref{eqn:Bethe}), or ``failure''.}
\State{For each $R\in\mathcal{R}$ and $j\in\partial R$, initialize $F_{j\to R}(z_j) \propto 1$.}
\While{neither converged nor timed-out}
	\State{Compute $V^{(R)}_i(z_i) = -\sum_{S\ni i \::\: S\not=R} T \log F_{i\to S}(z_i)$}
	\State{Obtain $b_R(\vec{z}^{(R)})$ by minimizing Eq.~\eqref{eq:regionalFreeEnergy} using the corrected energy $\tilde{E}_R$}
	\State{Compute the messages $F_{R\to i}(z_i) \propto \left[\sum_{\vec{z}^{(R)}\setminus z_i} b_R(\vec{z}^{(R)})\right]/F_{i\to R}(z_i)$.}
	\State{Re-estimate $F_{i\to R}(z_i) \propto \prod_{S\ni i\::\: S\not=R} F_{S\to i}(z_i)$.}
	\EndWhile
\end{algorithmic}
\caption{Generalized belief propagation (GBP) based on regional decomposition.}\label{fig:gbp_alg}
\end{algorithm}

Algorithm \ref{fig:gbp_alg} is a generalized belief propagation (GBP) that uses the Boltzmann distribution of each region's corrected penalty model to re-estimate their collective corrective biases. If this algorithm converges, then one obtains a critical point of the regional Bethe approximation (\ref{eqn:Bethe}) constrained to give consistent marginals $\sum_{\vec{z}^{(R)}\setminus z_i} b_R(\vec{z}^{(R)}) = b_i(z_i)$, \cite{Lackey15}. Like belief propagation, there is generally no guarantee of convergence and standard relaxation techniques, such as bounding messages away from $0$ and $1$, are required.

Beyond a proof of correctness, GBP offers a distinct computational advantage over our previous belief propagation algorithm from \cite{Bian14}. For ease of reference, we include the relevant message formulation from that work:
\begin{equation*}\label{eqn:bp1}
\mu_{R \rightarrow i}(z_i) := 
\min_{\vec{z}\backslash z_i} \left\{ \sum_{(j,k) \in E^{(R)}} J^{(R)}_{j,k} z_j z_k + \sum_{i \in V^{(R)}} h^{(R)}_i z_i + \sum_{j \in \partial R\backslash i} \mu_{j \rightarrow R}(s_j) \right\}.
\end{equation*}
Note there are $2 |\partial R|$ Ising model energy minimizations to be performed in each region $R$. With current QA hardware, programming of $\vec{h},\vec{J}$ parameters is significantly slower than sampling many solutions, and thus the cost of $2 |\partial R|$ reprogrammings can be significant. In GBP however we use QA hardware not to estimate a ground state energy, but to approximate the distribution $b_R(\vec{z}^{(R)})$. This can be performed with a single programming call per region. Each message is formed from the marginals, $\sum_{\vec{z}^{(R)}\setminus z_i} b_R(\vec{z}^{(R)})$, which are estimated from the hardware sampled ensemble.

One weakness in this algorithm is the need to know the temperature $T$ in order to produce the corrective biases $V^{(R)}_i(z_i)$. \cite{benedetti2015estimation} and \cite{Raymond16} propose methods to estimate instance dependent effective temperature directly from samples. It seems likely these techniques can be applied to GBP, and will be incorporated into future work.

\section{Application: fault diagnosis}\label{sec:faults}

We apply the methods of the previous sections to solve problems in fault diagnosis, a large research area supporting an annual workshop since 1989\footnote{\textit{E.g.} \url{http://dx15.sciencesconf.org/}}. We focus on circuit hardware fault diagnosis, which has featured as the ``synthetic track'' in four recent international competitions \cite{kurtoglu2009first,poll2011third}. Our goal is to use fault diagnosis as an example of how to use the methods of this report, and we use these competitions as inspiration rather than adhere to their rules directly.  The typical problem scenario is to inject a small number of faults into the circuit, using the specified fault modes for the targeted gates, and produce a number of input-output pairs. Now, given only these input-output pairs as data, one wishes to diagnose the faulty gates that lead to these observations. As typically there will be many valid diagnoses, the problem is to produce one involving the fewest number of gates (a ``min-fault" diagnosis).

We restrict to the ``strong'' fault model, in which each gate is healthy, and behaves as intended, or fails in a specific way. (In the ``weak'' fault model only healthy behaviour is specified.) The strong fault model is generally considered more difficult that then weak model, but is no harder to describe using the Ising model techniques of \S{\ref{sec:constraints}}.

Both the strong and weak fault model diagnosis problems are NP-hard. State-of-the-art performance for deterministic diagnosis is achieved by translating the problem into a SAT instance and using a SAT solver \cite{Metodi14}, but this approach has not been as thoroughly investigated in the strong fault model \cite{Stern14}. Greedy stochastic search produces excellent results in the weak fault model, but is less successful in the strong fault model \cite{Feldman10}.

We study the effectiveness of the D-Wave hardware in two experiments. First, we examine the ability of the hardware to sample diverse solutions to a problem. We find, despite not sampling diagnoses uniformly, that almost all min-cardinality diagnoses can be produced by oversampling the hardware by a factor of $1000$ (Table \ref{table:74XXXresults}). Next, we use the hardware to produce a solution for a problem too large to be embedded. We test dual decomposition from \cite{Bian14} and divide-and-concur from \S{\ref{sec:decomposition}} above, and solve several min-fault diagnosis problems that require multiple regions (Figure~\ref{figure:ISCASperf1}).


\subsection{Problem generation}\label{sec:prob_generation}

We test on the ISCAS '85 benchmarks \cite{Hansen99} and 74X-Series combinatorial logic circuits. From publicly available .isc files\footnote{ \url{http://web.eecs.umich.edu/~jhayes/iscas.restore/benchmark.html}}, we remove fault modes for buffer or fan-out wires, leaving only fault models for gates. Additionally, in order to accommodate penalty modeling with a small number of variables, we split certain large gates into smaller ones; this can be done without changing the correct fault diagnoses.

Owing to the difficulty of generating good input-output pairs \cite[\S{4.2}]{poll2011third}, we take a simplified approach. For each circuit, we randomly generate 100 observations (input-output pairs) and select a subset of size 20 with as uniform a distribution of minimum fault cardinalities as possible. These cardinalities are verified using a the MAX-SAT solver EVA \cite{Narodytska14}.

We perform cone-clustering (\S{\ref{sec:clustering}}) on each circuit using the ``pessimistic'' approach for strong-fault models of \cite{Stern14}, and generate Ising models to represent the constraints for each cone. When using explicit health variables, the resulting Ising models have energy gap at least $2$, while with implicit health variables, the energy gap is $1$ (using hardware-structured Ising models with $J_{ij} \in [-1,1]$ and $h_i \in [-2,2]$).

We partition the cone-clusters into regions using the software package METIS \cite{Karypis98}, with the number of regions chosen so that each region is embeddable in a working D-Wave hardware graph with up to $1152$ qubits. Finally we embed each region using the ``rip-up-and-replace'' algorithm of \S{\ref{subsubsec:rrr}}. It is important to note that for a given circuit, each of its regions need only be embedded once as different test observations may use the same embedding. Table~\ref{table:ISCASemb} summarizes the circuits, partitions, and embedding statistics.

\begin{table}[t]
\renewcommand{\arraystretch}{1.2}
\begin{center}
\begin{tabular}{@{}lllccllccll@{}}
\toprule 
& & & & \multicolumn{3}{c}{Explicit faults} & & \multicolumn{3}{c}{Implicit faults} \\
\cmidrule{5-7} \cmidrule{9-11}
Name & Gates & Var's & & $|\mathcal{R}|$ & \parbox{25pt}{Qubits/\\ region} & \parbox{20pt}{Chain length} & & $|\mathcal{R}|$ & \parbox{25pt}{Qubits/\\ region} & \parbox{20pt}{Chain length} \\
\midrule
74182 & 18 & 27 & & 1 & 241 & 8 & & 1 & 197 & 8 \\
74L85 & 25 & 36 & & 1 & 376 & 12 & & 1 & 315 & 12 \\
74283 & 30 & 39 & & 1 & 430 & 17 & & 1 & 329 & 15 \\ \hline 
c432 & 124 & 160 & & 3 & 395-499 & 22-35 & & 2 & 561-563 & 33-35 \\
c499 & 162 & 203 & & 4 & 416-460 & 18-31 & & 3 & 411-439 & 8-31 \\
c880 & 287 & 347 & & 4 & 496-635 & 8-22 & & 3 & 544-574 & 13-15 \\
c1355 & 474 & 515 & & 6 & 486-639 & 10-32 & & 4 & 506-553 & 8-34 \\
c1908 & 379 & 412 & & 7 & 534-684 & 18-39 & & 5 & 584-763 & 13-44 \\
\bottomrule
\end{tabular}
\end{center}
\caption{Statistics for the 74X Series and ISCAS '85 benchmarks as embedded on a D-Wave 2X processor, including number of regions $|\mathcal{R}|$ in the decomposition. Chain length refers to the maximum size of a chain within each region.}
\label{table:ISCASemb}
\end{table}

\subsection{Generating diverse solutions}\label{sec:sampling}

To test the D-Wave hardware's ability to generate diverse solutions, we consider the problem of finding \emph{all} min-cardinality fault diagnoses for a given observation. This is computationally more difficult than finding a single diagnosis, but also more realistic from the perspective of applications. Again, state-of-the-art performance in the weak fault model is achieved using a SAT-solver \cite{Metodi14}.

The hardware's natural ability to rapidly generate low-energy samples lends itself to applications in which a diverse set of optimal solutions are required. Unfortunately, samples taken from the D-Wave hardware do not conform to a Boltzmann distribution, owing to both noise and quantum mechanical effects. In contrast with greedy stochastic search \cite{Feldman10}, min-cardinality solutions will not generally be sampled with equal probability. In practice, Gibbs sampling \cite{Geman84} and other post-processing techniques may be used to make a distribution of ground states more uniform.

We restricted to the 74X-Series circuits in Table \ref{table:ISCASemb}, which can entirely embedded within the current hardware architecture. For each input-output pair for a circuit, we used SharpSAT \cite{Thurley06} to enumerate the min-cardinality diagnosis set $\Omega^{\leq}$ and then drew $1000|\Omega^{\leq}|$ samples from the QA hardware. Ising models were pre-processed with roof-duality \cite{Boros02} and arc-consistency \cite{Mackworth77}, allowing certain variables to be fixed in polynomial time. Random spin-reversal transformations (``gauge transformations'') were applied to mitigate the effects of intrinsic control error in the D-Wave hardware. Samples were post-processed using majority vote to repair broken chains, followed by greedy bit-flipping in the original constraint satisfaction space to descend to local minima. See \cite{King14} for more details on pre- and post-processing.

\begin{figure}
\centering
\parbox{0.5\textwidth}{
  \includegraphics[width=3.0in]{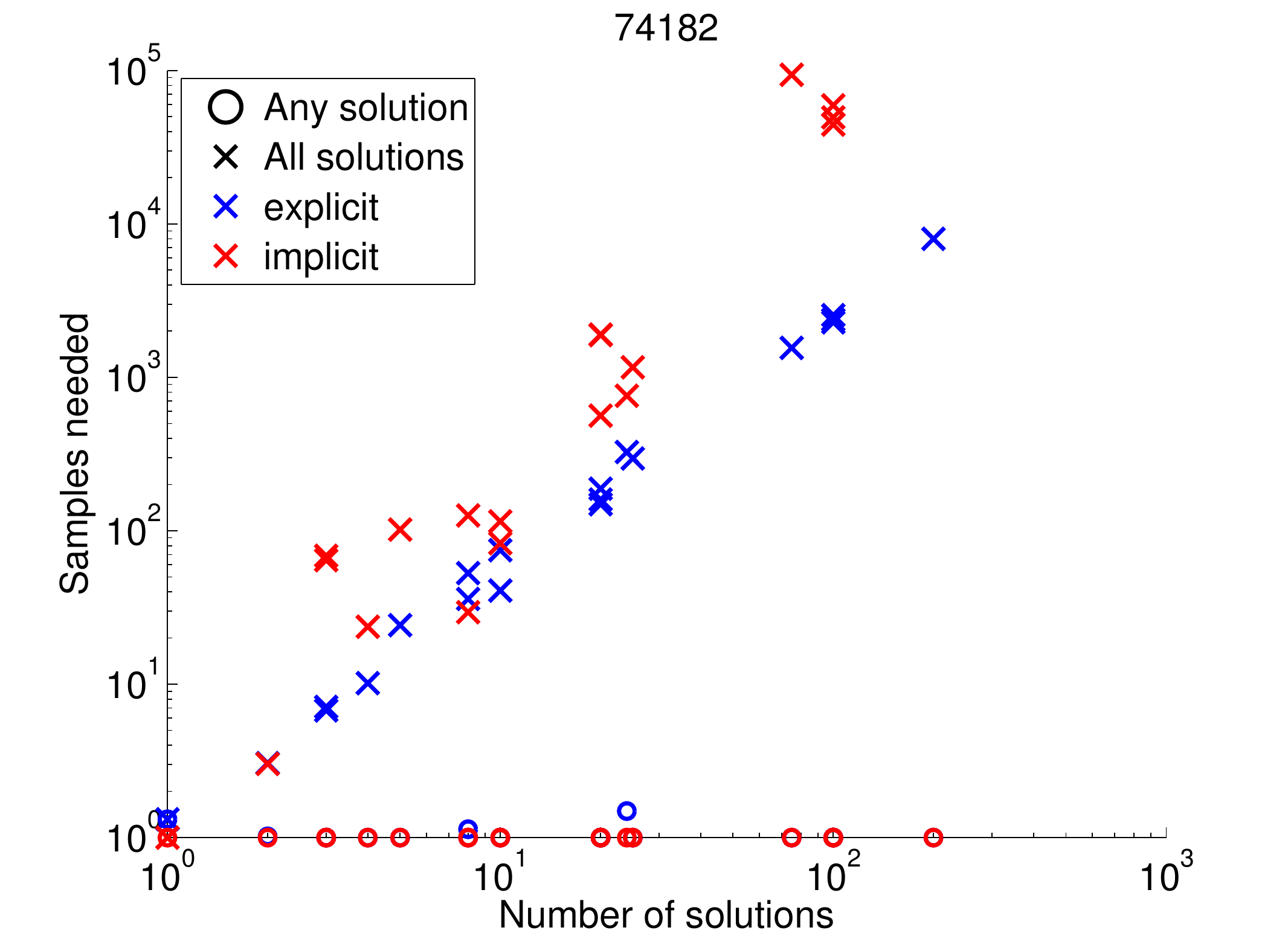}
  }%
\parbox{0.5\textwidth}{
  \includegraphics[width=3.0in]{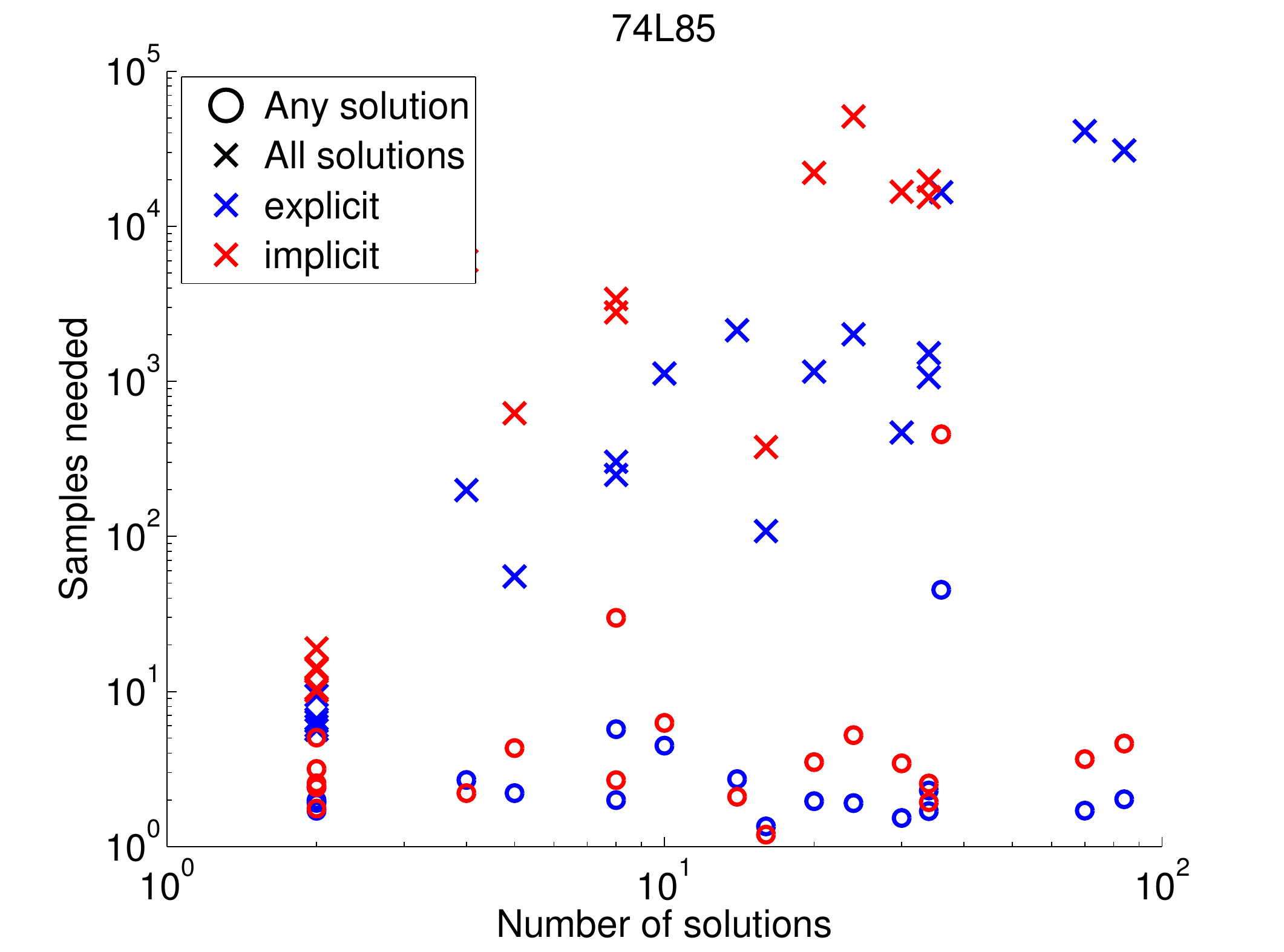}
  }%
\\
\parbox{0.5\textwidth}{
  \includegraphics[width=3.0in]{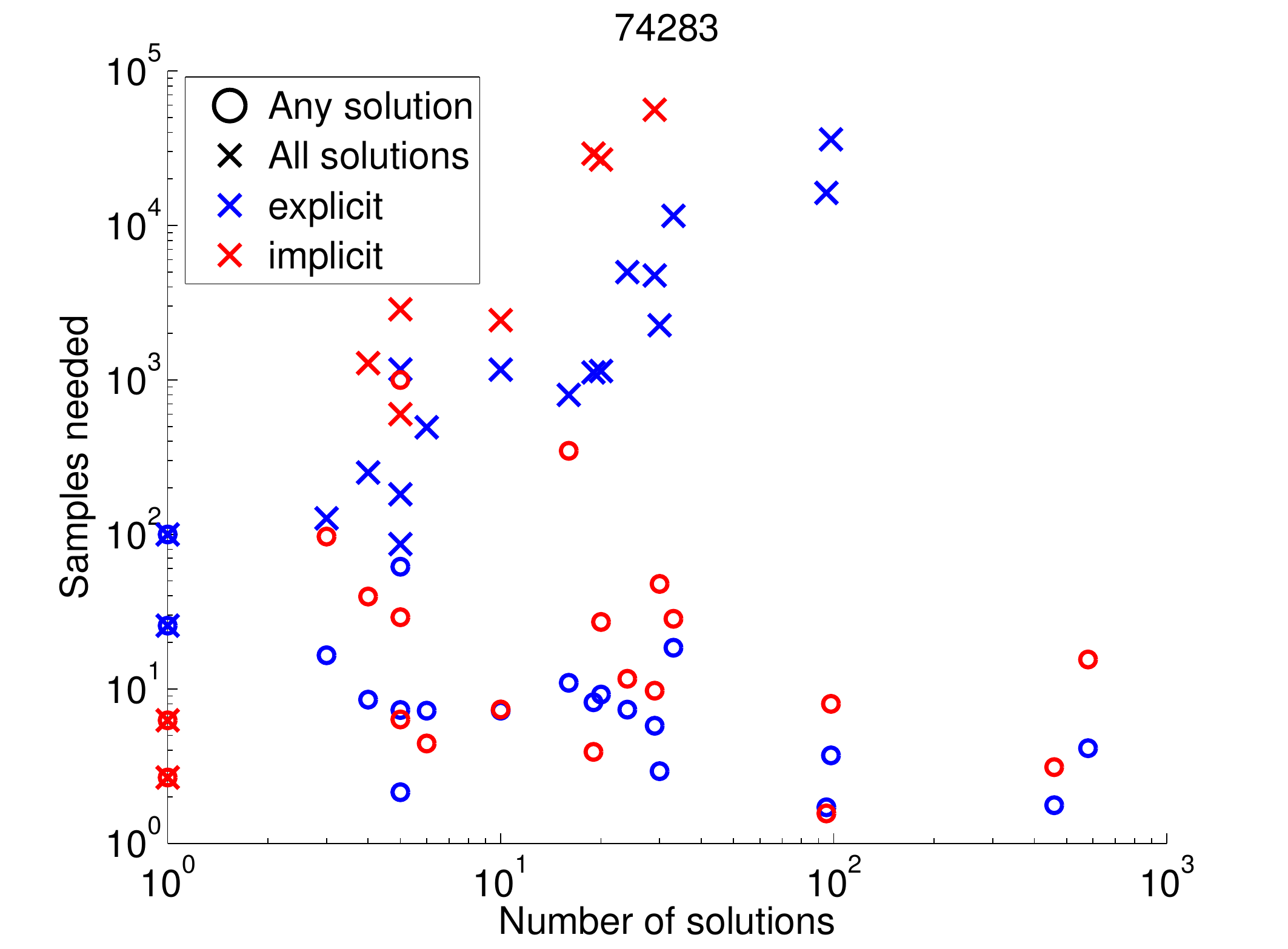}
  }%
  \caption{Performance in finding all min-fault diagnoses for the 74X
    benchmarks using a D-Wave 2X processor.  Missing $\times$'s indicate that not all solutions were found.}
  \label{figure:74XXXresults}
\end{figure}

The results in Figure~\ref{figure:74XXXresults} show the expected number of samples needed to see all min-fault diagnoses at least once, together with the number of samples needed to see just a single min-fault diagnosis. Namely, if $p_i$ denotes the fraction of all samples taken that correspond to min-fault diagnosis $i$, then the expected number of samples required to find a single min-fault solution is $1/\sum_i p_i$, and the expected number of samples required to find all min-fault solutions is $\int_0^{\infty} (1-\prod_i(1-e^{-p_it}))\; dt$. (This is the coupon collecting problem with non-uniform probabilities \cite{Schelling54,Flajolet92}.) Following \cite{Feldman10}, we also computed the expected fraction of all min-fault diagnoses found when taking $N|\Omega^{\leq}|$ samples, for $N \in \{10,100,1000\}$. These results are summarized in Table \ref{table:74XXXresults}.

\begin{table}
\begin{center}
\begin{tabular}{@{}llclllclll@{}} 
\toprule
  & & & \multicolumn{3}{c}{Explicit faults} & & \multicolumn{3}{c}{Implicit faults} \\
\cmidrule{4-6} \cmidrule{8-10}
  Name & $|\Omega^{\leq}|$ & & $M_c(10)$ & $M_c(100)$ & $M_c(1000)$ & & $M_c(10)$ & $M_c(100)$ & $M_c(1000)$ \\
  \midrule
  74182 & 1-200 & & 95.5 & 100 & 100 & & 63.9.8 &  90.0 & 98.9 \\
  74L85 & 2-84 & & 69.5 & 94.9 & 100 & & 44.7 &  71.0 & 90.1\\
  74283 & 1-580 & & 60.4 & 91.2 & 98.6 & & 25.9 &  56.2 & 80.0\\
\bottomrule
\end{tabular}
\end{center}
\caption{Performance in finding all min-fault diagnoses for the 74X benchmarks using a D-Wave 2X processor. $\Omega^{\leq}$ is the set of min-fault diagnoses for a given instance, and $M_c(N)$ is the expected percentage of all min-fault diagnoses found when $N|\Omega^{\leq}|$ samples are taken for each instance. Note that the annealing time to take $100$ samples is 2 ms, roughly the same as the time to take $4$ samples reported in Table 6 of \cite{Feldman10}.}
\label{table:74XXXresults}
\end{table}

\subsection{Solving large problems}

We tested the performance of the D-Wave hardware in solving the fault diagnosis problem for circuits too large to be embedded. On the regions produced in \S{\ref{sec:prob_generation}}, we applied two algorithms: dual decomposition (DD) from \cite{Bian14} and divide-and-concur (DC) from \S{\ref{sec:decomposition}}. To measure algorithm performance independent of quantum annealing, we also found the minima for regional Ising models exactly using low-treewidth variable elimination \cite{Koller09}. Such an exact solver (SW) gives an upper-bound on the performance of a decomposition algorithm.

\begin{figure}
\centering
\begin{picture}(450,155)
	\put(0,145){(A)}
	\put(25,0){\includegraphics[width=2.5in]{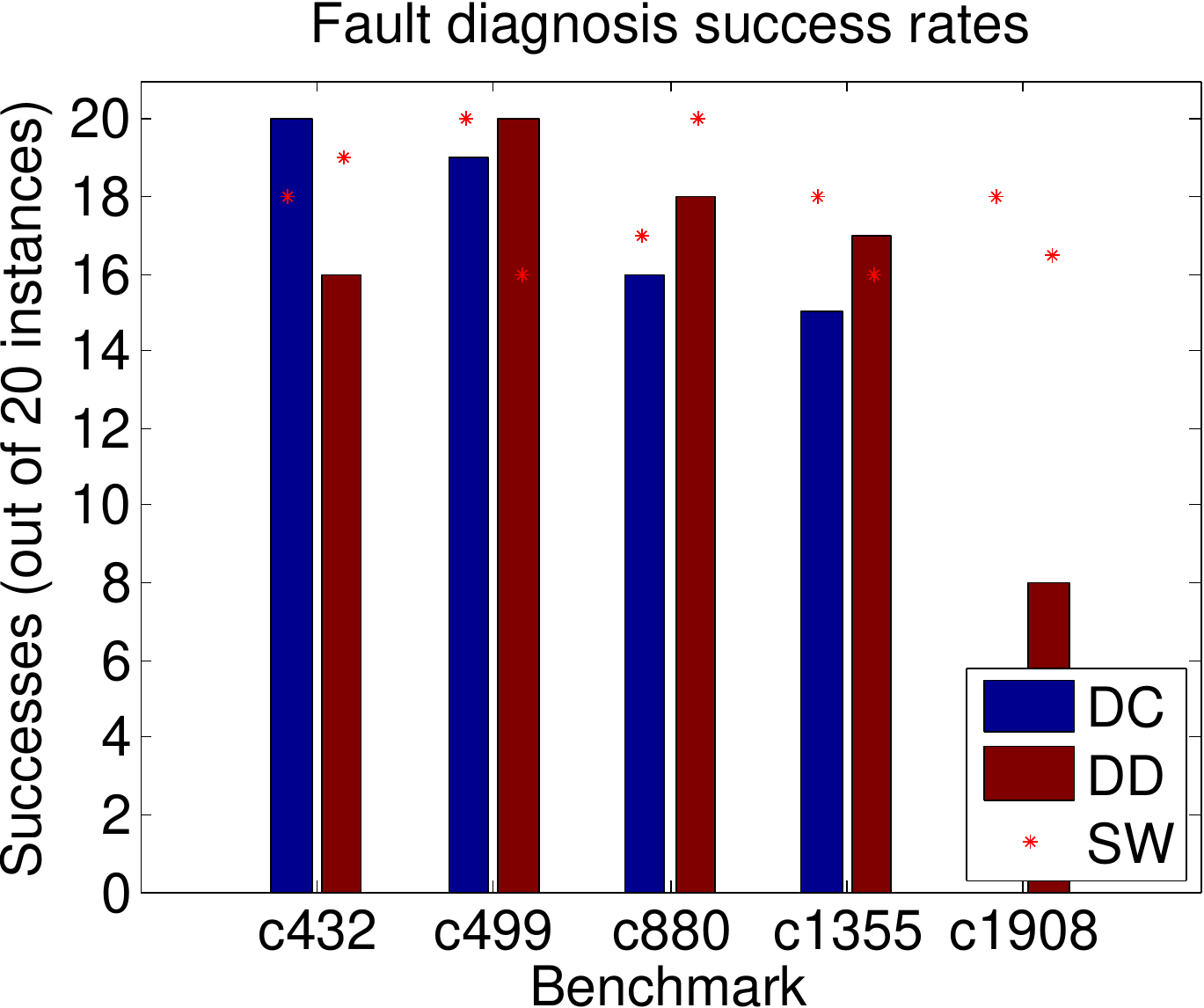}}
	\put(225,145){(B)}
	\put(250,0){\includegraphics[width=2.5in]{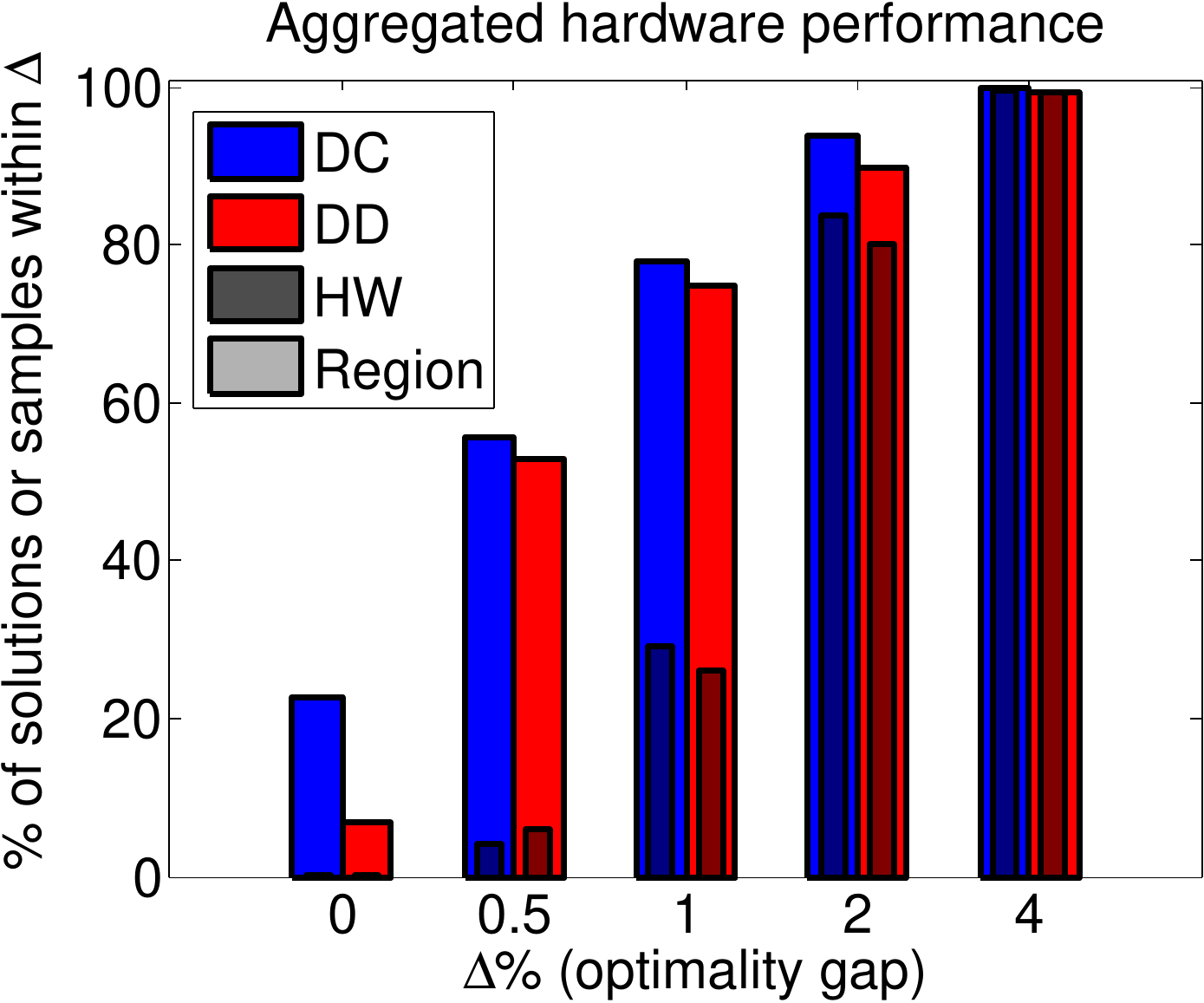}}
\end{picture}
\caption{(A) Summary of D-Wave hardware success rate using divide-and-concur (DC) and dual decomposition (DD), compared to the same decomposition algorithm using an exact low-treewidth software solver (SW). (B) Percentage of hardware samples (HW) and regional solutions (Region) within $\Delta\%$ of optimality across all instances tested.}
\label{figure:ISCASperf1}
\end{figure}

In Figure~\ref{figure:ISCASperf1}(A), we show the number of successful min-fault diagnoses out of $20$ instances for several of the ISCAS '85 benchmark circuits. Using the exact solver we attempted to solve each fault-diagnosis instance $100$ times, and recorded the median number of successes across the $20$ instances for each circuit. Using the D-Wave hardware, we attempted to solve each instance once. A summary of the D-Wave hardware performance on each regional optimization problem is given in Figure~\ref{figure:ISCASperf1}(B).  Each problem was solved by drawing 20,000 samples across 20 spin reversal transformations, with pre- and post-processing as in \S{\ref{sec:sampling}}. 

Note that the overall performance of the decomposition algorithms using D-Wave's heuristic optimizer is similar to that using an exact solver, despite the fact that the D-Wave hardware does not solve every sub-problem to optimality. This suggests that QA hardware that provides only approximate solutions in the form of low-energy samples can still be used to solve large optimization problems, provided it can capture a sufficiently non-trivial portion of the original problem.

\section{Conclusion}\label{sec:conclusion}
In this paper we have expanded on the approach given in \cite{Bian14}
to solve large discrete optimization problems using quantum annealing
hardware limited by issues of precision, connectivity and
size. Applying these techniques with the D-Wave 2X device, we are able
to solve non-trivial problems in model-based fault
diagnosis. While the total running times of the decomposition algorithms are not currently competitive with the fastest classical techniques, both the speed and the performance of the algorithms improve dramatically with the size of the quantum hardware available.

Two of the most important directions for future research are:
\begin{enumerate}
\item {\bf Expanding penalty-modelling techniques to more qubits.} As
  the available hardware grows larger, large energy gaps and other
  forms of error-correction will become more important to finding the
  grounds state in quantum annealing. In addition, a better
  understanding of the performance trade-off between larger energy gap
  and fewer qubits is needed.
\item {\bf Alternate strategies for decomposition algorithms.} Since
  minor-embedding is itself a difficult discrete optimization problem,
  current decomposition algorithms are hampered by the need for fixed
  regions with pre-computed embeddings. More research is needed into
  circumventing the need for fixed regions, combining quantum annealing with the
  best classical constraint satisfaction methods, and making better use of the fast sampling capabilities of the available hardware.
\end{enumerate}

\bibliographystyle{plain}

\end{document}